\documentclass[12pt]{amsart}
\usepackage{amssymb,amsmath,latexsym} 
\def\cal{\mathcal}
\newtheorem{theorem}{Theorem}

\newtheorem{exe}[theorem]{Exercise}
\newtheorem{exa}[theorem]{Example}
\newtheorem{defi}[theorem]{Definition}
\newtheorem{remark}[theorem]{Remark}
\newenvironment{rem}{\begin{remark} \rm}{\end{remark}}

\newcommand{\CD}{{\cal D}}

\makeatletter
\@addtoreset{equation}{section}

\makeatother

\def\be{\beta}
\def\al{\alpha}

\def\la{\lambda}

\newcommand{\rref}[1]{(\ref{#1})} 

\newcommand{\del}{{\partial}}
\newcommand{\som}{{\sum_{i=1}^{n+1}}}

\def\Nij{Nijenhuis}

\def\bih{bi-Ham\-il\-tonian}

\def\dncoo{Darboux-\Nij\ coordinates}
\def\omnman{$\omega N$ manifold}
\def\on{$\omega N$\ }
\def\pd#1{\frac{\partial}{\partial#1}}
\def\diag{\operatorname{diag}}
\begin{document}
\title[Separability of the NR system]{A geometric approach to the separability
of the Neumann-Rosochatius system}
\author[C. Bartocci]{Claudio Bartocci}\author[G. Falqui]{Gregorio Falqui}
\author[M. Pedroni]{Marco Pedroni}
{ {\renewcommand{\thefootnote}{\fnsymbol{footnote}}
\footnotetext{\kern-15.3pt {\em AMS Subject Classification:} 37K10, 53D17,70H20}
}}
{ {\renewcommand{\thefootnote}{\fnsymbol{footnote}}
\footnotetext{\kern-15.3pt{\em Keywords:} Bi-Hamiltonian manifolds, Separation of variables, Spectral Curves}
}}
{ {\renewcommand{\thefootnote}{\fnsymbol{footnote}}
\footnotetext{\kern-15.3pt
SISSA Preprint 62/2003/FM 
}}
\begin{abstract} \noindent
We study the separability of the Neumann-Rosochatius system on the
$n$-dimensional sphere using the geometry of bi-Hamiltonian manifolds.
Its well-known separation variables are recovered by means of a
separability condition relating the Hamiltonian with a suitable
$(1,1)$ tensor field on the sphere. This also allows us to iteratively
construct the integrals of motion of the system.
\end{abstract}
\maketitle
\section{Introduction}
The Neumann system 
is among the widest known and best studied integrable
systems in Mathematical Physics. It decribes the dynamics of a
point particle constrained to move on the sphere $S^n$, under the
influence of a quadratic potential $V(x)=\frac12
\sum_{i=1}^{n+1}\al_i x_i^2, \>
\al_i\neq\al_j$.
In 1859, Carl Neumann \cite{neumann} showed that the
equations of motion of the ``physical'' $n=2$ case could be solved using the
Jacobi theory of separation of variables.
It was noticed by Rosochatius (see \cite{ratiu})
that a potential given
by the sum (with nonnegative weights) of the inverses of the squares
of the (Cartesian)
coordinates can be added without losing the separability property.
The system so obtained is customarily called the {\em
Neumann-Rosochatius} (NR) system.

More than one century later this separability
result was generalized to the arbitrary $n$ case by Moser \cite{Moser}.
The starting point to solve the problem was the ingenious introduction of a
special set of coordinates on $S^n$, called spheroconical (or
elliptical spherical)
coordinates (already used, for $n=2$, by Neumann).
They are defined as follows: For given sets of real numbers
$\al_1<\al_1<\cdots<\al_{n+1}$ and nonzero
$x_1,\ldots, x_{n+1}$, the coordinates $\la_a(x),
a=1,\ldots,n$,
are the solutions of the equation
\begin{equation}
      \label{eq:i1}
      \sum_{i=1}^{n+1}
\frac{x_i^2}{\la-\al_i}=0\ .
\end{equation}
Later on, it was shown that the NR system could be framed within the
formalism of Lax pairs and $r$-matrices (see,
e.g.,~\cite{AHH93,Harnadlectures}).
Actually, it turns out  that introducing the Lax matrix,
as a function of the Cartesian coordinates
$x_i,y_i,\>i=1,\ldots, n+1$, by
\begin{equation}\label{eq:i2}
N(\la)=\left(
\begin{array}{cc} -h(\la)+ik(\la)&e(\la)\\ f(\la)&
h(\la)+ik(\la)\end{array}\right)\ ,
\end{equation}
where
\begin{equation}\label{eq:i3}
\begin{split} &h(\la)=\frac12\sum_{i=1}^{n=1} \frac{x_i y_i}{\la-\al_i},\quad
k(\la)=\frac12\sum_{i=1}^{n=1} \frac{\beta_i}{\la-\al_i},\\
&f(\la)=\frac12\sum_{i=1}^{n=1} \frac{x_i^2}{\la-\al_i},\qquad
e(\la)=-\frac12\left(1+\sum_{i=1}^{n=1}
\frac{y_i^2+\beta_i^2/x_i^2}{\la-\al_i}\right),
\end{split}
\end{equation}
and identifying the cotangent bundle to $S^n$ with the submanifold
of ${\mathbb R}^{2(n+1)}$ defined by the constraints
\begin{equation}\label{eq:i3a}
\som x_i^2=1,\qquad \som x_iy_i=0\ ,
\end{equation}
the Hamilton equations of motion of the NR system acquire the form
\begin{equation}\label{eq:i4}
\frac{d N(\la)}{dt}=[\Phi,N(\la)],
\end{equation}
where
\[
\Phi=\left(\begin{array}{cc}
\som x_iy_i&\la+\som ({y_i}^2+\al_i {x_i}^2\beta_i^2/x_i^2)\\
-\som x_i^2&-\som x_iy_i
\end{array}\right).
\]
As a consequence, the spectral invariants of $N(\la)$ are constants
of motion.
In particular, the quantities
\[
K_i=\text{res}_{\big\vert_{\la=\alpha_i}}\text{det}(N(\la))\ ,
\qquad i=1,\dots,n+1
\]
(known as Uhlenbeck integrals), provide $n$ mutually commuting integrals of
motion that ensure Liouville--Arnol'd integrability of the NR system, the
physical Hamiltonian being given by
\[
H_{NR}=2\sum_{i=1}^{n+1} \alpha_i K_i+\frac12\sum_{i>j=1}^{n+1} \be_i\be_j.
\]
Separation of variables is recovered in this formalism noticing that,
on $S^n$, the
zeroes $\{\la_a\}_{a=1,\ldots,n}$
of the matrix entry $f(\la)$ define the spheroconical coordinates, and their
conjugate momenta are given (as it will be explicitly recalled in
Section \ref{sec:3}) by
the values of the rational function $h(\la)$ for $\la=\la_a$.
Clearly, each pair
of canonical coordinates $(\la_a,\mu_a)$ satisfy the separated equation
\begin{equation}\label{eq:i6}
\text{det}(N(\la_a))+\mu_a^2+k^2(\la_a)=0, \quad a=1,\ldots,n.
\end{equation}
In this paper we want to provide a further geometrical
interpretation of
the NR system, based on the notions of \bih\ geometry, generalizing and
refining the approach described in \cite{marconeu}.
We will follow a recently introduced set up for the theory of separation of
variables for the Hamilton-Jacobi equations.
In a nutshell, such a framework can be described as follows. One considers
a symplectic manifold $M$ endowed with a  $(1,1)$ tensor field $N$ with
vanishing \Nij\ torsion (which we will call an {\em
   \omnman\/}, provided that a compatibility condition between $N$ and
the symplectic form is satisfied);
under suitable hypotheses, $N$ selects
a special subclass of canonical coordinates on $M$ (called \dncoo)
that have the
property of diagonalizing $N$. The condition for the
separability of the Hamilton-Jacobi equation associated with a Hamiltonian
$H$ can be given, according to the \bih\ theory of separation of variables, the
following intrinsic formulation. One considers the distribution
$\CD_H$ generated
iteratively by the action of $N$ on the Hamiltonian vector field
$X_H$ associated
with $H$, and the two--form $d(N^* dH)$. Then $X_H$ is
separable in the
\dncoo\ associated with $N$ if and only if
\begin{equation}\label{eq:i7}
     d(N^* dH)(\CD_H,\CD_H)=0.
\end{equation}
This scheme, in its basic features, 
has already been considered in the literature \cite{bcrr96,mt97,fmt,cetraro,FP}
and applied to various systems 
(see, e.g, \cite{FMPZ2,creta,Bl98,mt02});
it is fair to say that, in these papers, the
\omnman\ structure is fixed ``a priori'', and that equation~\rref{eq:i7}
is seen as a condition that
selects those Hamiltonians which are separable in the ``preassigned'' \dncoo.

In the present paper we will take a different logical standpoint: We
will consider a given
Hamiltonian $H$ (namely, the NR Hamiltonian) and look
at~\rref{eq:i7} as an equation to determine $N$ (and hence the separation
coordinates). We shall see that it is indeed possible (and, actually,
easy) to solve
such an equation by means of a couple of natural {\em Ans\"atze}, thus
arriving to
induce from $H$ the separation coordinates. Also, we shall show
how the iterative
structure naturally associated with the (generalized) recursion
relations defined by $N$
allow to recursively construct the additional integrals of motion ensuring
complete integrability. Finally, we will make contact with the ``Lax'' approach
to the separability of the Neumann-Rosochatius system showing that the
separation relations tying \dncoo\ and these integrals are nothing but
the spectral curve relations~\rref{eq:i6}.

Obviously enough, the conditions on $N$ coming from~\rref{eq:i7},
in their full generality, are too
difficult to be solved. The couple of Ans\"atze which will enable us to solve
them for the NR case are the following. The first one is suggested from the
fact that the phase space of the NR system is a cotangent bundle; accordingly,
we will seek for a special \omnman\ structure on $T^* S^n$, defined
by a $(1,1)$
tensor $N$ induced by a suitable tensor (with zero torsion) $L$ on the base
manifold $S^n$. The second one will be to use a special form of
equation~\rref{eq:i7}, that reads
\begin{equation}\label{eq:i8}
d\big(N^*dH-\frac12\text{tr}(N)\wedge dH\big)=0.
\end{equation}
The plan of the paper is as follows: In Section \ref{sec:1} we will collect
some notions of the theory of \omnman s, and briefly discuss the
\bih\ theorem for separation of variables.
In Section \ref{sec:2} we will solve equations~\rref{eq:i8}, thus
showing that the
geometry of $\omega N$ manifolds can be used to discover the separation
variables of the Neumann-Rosochatius (NR) system. In Section
\ref{sec:3} we will
find the St\"ackel separation relations and the family of commuting
integrals of the system.
\par\smallskip\noindent
{\bf Acknowledgments.}
We wish to thank Sergio Benenti, Franco Magri, and Giovanni Rastelli
for useful discussions. The results presented in Section 2 have been
obtained in collaboration with Franco Magri.
This work has been partially supported by the
Italian M.I.U.R. under the research project {\em Geometry of
Integrable Systems.}

\section{The $\omega N$ framework}\label{sec:1}

In this section we wish to recall some basic properties of a special
class of bi-Hamil\-to\-nian manifolds, called \on
manifolds. For a more detailed description we refer to \cite{fratiz,FP}.
By definition, an \on manifold is a smooth manifold $M$ endowed
with a pair of compatible Poisson bivectors $P_0$, $P_1$ such that
one of them (say, $P_0$) is nondegenerate. (Compatibility means
that any linear combination of $P_0$ and $P_1$ is a Poisson bivector.)
One can construct a recursion operator $N= P_1 P_0^{-1}$, whose
Nijenhius torsion,
\begin{equation}
T(N) (X,Y) = [NX,NY] - N([NX,Y]+ [X,NY] - N[X,Y])\ ,
\end{equation}
vanishes as a straightforward consequence of the compatibility between
$P_0$ and  $P_1$ (see, e.g., \cite{magri8}).
We set $2n = \hbox{dim}\, M$ and we denote by
$\omega_0$ the symplectic structure associated to
$P_0$, and by $\{\cdot, \cdot\}_0$,
$\{\cdot, \cdot\}_1$ the Poisson brackets associated, respectively,
to $P_0$, $P_1$.

The relevance of \on manifolds in the theory of separable systems is mainly
due to the existence, under
suitable hypotheses, of a special class of canonical coordinates,
that are selected by the geometric structure of the system itself.
\begin{defi} A system of local coordinates
$(x_1,\dots,x_n,y_1,\dots,y_n)$ that are canonical w.r.t.\ the
symplectic form
$\omega_0$ is said to be
Darboux-Nijenhuis if the matrix expressing $N$ in these coordinates is
diagonal, i.e.,
$$ N = \sum_{i=1}^{n} \left(\lambda_i \pd{x_i} \otimes dx_i + \nu_i
\pd{y_i} \otimes dy_i\right)\,.
$$
\end{defi}
Notice that, since $NP_0$ is antisymmetric, it follows that $\lambda_i
=\nu_i$ for all $i$. In general, however,
the eigenvalues $\lambda_1,\dots, \lambda_n$ need not be distinct.

On the cotangent bundle of any differentiable manifold there is an
elegant way to construct
\on structures admitting Darboux-Nijenhuis coordinates
through the following procedure \cite{IMM}.
Let $Q$ be an $n$-dimensional manifold equipped with a type
$(1,1)$ tensor field $L$, whose Nijenhuis
torsion vanishes. Let $\theta_0$ be the Liouville
$1$-form and $\omega_0= d\theta_0$ the standard symplectic
$2$-form on $T^\ast Q$; the associated Poisson structure will be denoted $P_0$.
By thinking of $L$ as an endomorphism of $TQ$, one can deform the
Liouville $1$-form to a $1$-form
$\theta_L$:
$$\langle \theta_L, Z\rangle_\alpha= \langle \alpha, L(\pi_\ast
Z)\rangle_{\pi(\alpha)} \,,$$
for any vector field $Z$ on $T^\ast Q$ and for any $1$-form $\alpha$ on
$Q$, where $\pi: T^\ast Q \to Q$ is the canonical
projection. If we choose local coordinates $(x_1,\dots,x_n)$ on $Q$ and
set $L(X) = \sum_{i,j=1}^n L^i_j X^j \pd{x_i}$, we get the local expression
$\theta_L = \sum_{i,j=1}^n L^i_j y_i dx_j$ w.r.t.\ the standard
symplectic coordinates $(x_1,\dots,x_n,y_1,\dots,y_n)$ on
$T^\ast Q$.
The {\it complete lift} of $L$ is the endomorphism $N$ of $T(T^\ast
Q)$ uniquely determined by the
condition
\begin{equation}
\label{defdiN}
d\theta_L (X,Y) = \omega_0(NX,Y)\,,
\end{equation}
for all vector fields $X$, $Y$ on $T^\ast Q$. An easy computation shows that:
\begin{align}
N (\pd{x_k}) &= \sum_{i}^n L^i_k \pd{x_i} -
\sum_{i}^n  y_l(\frac{\partial L^l_i}{\partial x_k} - \frac{\partial
L^l_k}{\partial x_i}) \pd{y_i}\\
N (\pd{y_k}) &= \sum_{i}^n L^k_i \pd{y_i}\ .
\end{align}
Since $L$ has vanishing
Nijenhuis torsion, the same property holds for the
type $(1,1)$ tensor field $N$ on $T^\ast Q$ \cite[Prop. 5.6, p.
36]{YanoIshihara}, and $(T^\ast Q,P_0,P_1:=NP_0)$ is
an \on manifold \cite{IMM}. The Poisson structure $P_1$ is related to
$\omega_1:=d\theta_L$ by the formula:
$$P_1 (dF,dG) = \omega_1(X_F,X_G)\qquad\hbox{for all}\ F,G\in
C^\infty(T^\ast Q)\,,$$
where $X_F$, $X_G$ are the Hamiltonian vector fields associated to
$F$, $G$ w.r.t.\ the symplectic form
$\omega_0$. By the very definition, if $X^{(1)}_H$ is the Hamiltonian
vector field associated to $H$ w.r.t.\ $P_1$, then  $X^{(1)}_H =
N X_H$. Notice that, in
general, this vector field need not be Hamiltonian or even locally
Hamiltonian w.r.t.\ $\omega_0$.
Indeed, the $1$-form $N^\ast dH$ may fail to be closed (here $N^\ast$
is the adjoint of the endomorphism $N$), and one has
\begin{equation}\label{eqxcs}
L_{NX_H}\omega_0 = -d (N^\ast dH)= L_{X_H}\omega_1\,.
\end{equation}
In fact, from \rref{defdiN} it follows that
$(N^\ast dF)(Y) = - \omega_0(NX_H,Y) =  - \omega_1(X_H,Y)$.

Let us now assume that $L$ has $n$ functionally independent
eigenvalues $\lambda_1,\dots,\lambda_n$. Since $L$ is
torsionless, these eigenvalues determine local coordinates  on $Q$
satisfying the relations:
$$ L \pd{\lambda_i} = \lambda_i \pd{\lambda_i}\,.$$
We denote by $\mu_i$ the conjugate momentum to $\lambda_i$; clearly,
one has $N \pd{\mu_i} = \lambda_i \pd{\mu_i}$.
The coordinates $(\lambda_1,\dots,\lambda_n, \mu_i,\dots,\mu_n)$ are
Darboux-Nijenhuis coordinates for the \on  manifold $(T^\ast Q, P_0,
P_1)$.

We can exploit the geometric setting of \on manifolds in order to
find an intrinsic
separability condition for  a given
Hamiltonian function $H \in C^\infty(T^\ast Q)$.
Let us suppose that the vector fields $X_H$, $NX_H,\dots N^{n-1}X_H$
are pointwise linearly independent, so that they
generate an $n$-dimensional distribution $\CD_H$.
If  we compute
the conditions
\begin{equation}\label{condns1}
d(N^\ast dH)(N^iX_H,N^jX_H)= 0 \qquad\hbox{for all}\ i,j=0,1,\dots,n-1
\end{equation}
in the Darboux-Nijenhuis coordinates $(\lambda_1,\dots,\lambda_n,
\mu_i,\dots,\mu_n)$, we get a system of differential
equations equivalent to the Levi-Civita separability formulae
\cite[p.~208, eq.~(1.230)]{DKN}.

\begin{theorem}\label{thmsep} In the above hypotheses
and notations, the Darboux-Nijenhuis coordinates
associated with $L$ are separation variables for $H$ if and only if
the $2$-form $d(N^\ast dH)$ annihilates the distribution $\CD_H$.
\end{theorem}
The separability condition \rref{condns1} implies that the
distribution $\CD_H$ is integrable. So, there exist $n$
independent local functions $H_1,\dots H_n$ that are constant on the
leaves of $\CD_H$. The distribution being invariant under
the action of $N$, the same is true for
   the differential
ideal generated by the $H_i$'s, so that the following condition holds:
\begin{equation}
N^\ast dH_i= \sum_{k=1}^n F_{ik} dH_k\,,
\end{equation}
where $F_{ik}$ is a matrix with distinct eigenvalues. Moreover, since
$\CD_H$ is
Lagrangian with respect to both $\omega_0$ and $\omega_1$,   we have:
\begin{equation}
\{H_i,H_j\}_0 = \{H_i,H_j\}_1 = 0 \qquad\hbox{for all} \ i,j\,.
\end{equation}
It follows  that the functions $H_1,\dots H_n$ are separable in the
the Darboux-Nijenhuis coordinates singled out by $L$.

A particular case of this state of affairs is provided by the
following example, which will turn out to
be of great importance in the study of the NR system.
Let us consider the characteristic polynomial
\begin{equation}
\label{2.7b}
\det(\lambda I - L) = \lambda^n -c_1\lambda^{n-1} - c_2\lambda^{n-2}
-\cdots - c_n
\end{equation}
of the endomorphism $L$, and assume we are given a Hamiltonian $H$
satisfying the condition
\begin{equation}\label{quellaforte1}
d(N^\ast dH) = dc_1 \wedge dH\,.
\end{equation}
  From \rref{eqxcs} it follows at once that this equation is equivalent to
\begin{equation}\label{quellaforte2}
d(L_{X_H}\theta_L - H dc_1)= 0\,.
\end{equation}
The condition  \rref{quellaforte1} is a sufficient condition to the
separability of $H$ in the Darboux-Nijenhuis
coordinates associated to $N$, because it implies that the $2$-form
$d(N^\ast dH)$ annihilates the distribution
$\CD_H$ generated by the vector fields $X_H$, $NX_H,\dots N^{n-1}X_H$.
Moreover, it can be shown \cite{IMM} that,
choosing a local function $H_2$ such that
$d H_2 = N^\ast
dH - c_1 dH$, the $1$-form $N^\ast dH_2 - c_2 dH$ is again closed, so
that we can find a local potential
$H_3$. By iterating this
procedure, we end up with
$n$ independent local functions $H_1=H,\dots, H_n$ that are constant on the
leaves of $\CD_H$ and satisfy the conditions:
\begin{align}\label{recursconds}
d H_{i+1} &= N^\ast dH_i - c_i dH \qquad i=2,\dots,n-1\cr
0 &= N^\ast dH_n - c_n dH\,.
\end{align}
In this case the matrix $F$ has the form
\begin{equation*}
F = \begin{pmatrix} c_1 &1 &0   &\dots  &0  \cr
c_2 &0 &1 &\ddots &\vdots \cr
\vdots &\vdots &0  &\ddots &\vdots \cr
\vdots &\vdots &\vdots &\ddots & 1\cr
c_n &0  &0  &\dots & 0\cr
\end{pmatrix}
\end{equation*}
and the following condition is readily checked:
\begin{equation}
N^\ast dF = F dF\,.
\end{equation}
We set $F= S^{-1}\diag (\lambda_1,\dots, \lambda_n )\, S$.
Then, one has $S_{jk}=\la_j^{n-k}$ and, by virtue of \cite[Theorem 4.2]{FP},
one obtains the separability equations
$\sum_{k=1}^n H_k \lambda_j^{n-k} = U_j$.
Summing up, the functions $H_1=H,\dots H_n$
are proved to be St\"ackel separable
in the Darboux-Nijenhuis coordinates associated to $L$
(this means that the separation relations are affine in the $H_k$'s).

\section{Torsionless tensors and separability of the NR system}\label{sec:2}

According to the results of Section 2, to separate the NR system we seek a
tensor field $L$ of type $(1,1)$ on $S^n$ satisfying the ``strong''
separability condition \rref{quellaforte1}, i.e.,
\begin{equation}
\label{2}
d(L_{X_H}\theta_L-H dc_1)=0\ ,
\end{equation}
where $\theta_L=\sum_{a,b=1}^n L^a_b p_a dq_b$
(for any set of fibered coordinates
$(q_a,p_a)$) and $c_1=\mbox{tr} L$. We have seen that the eigenvalues of
such an $L$ (if real and functionally independent) are separation variables
for $H$.
The form of the constraint and of the potential suggest
using on $S^n$ the coordinates $X_a:={x_a}^2$,
for $a=1,\ldots,n$.\footnote{
In this and in the following section we use the following convention: middle
indices like $i,j,k$ run from $1$ to $n+1$, while indices like $a,b,c$ run from
$1$ to $n$.} If $Y_a$ are the momenta conjugated to the $X_a$ (and the point
particle has unit mass), the NR Hamiltonian is given by $H=T+V$ with
\begin{eqnarray*}
T &=& 2\sum_{a} X_a (1-X_a){Y_a}^2-4\sum_{a<b}X_a X_b Y_a Y_b\\
V &=&
\frac12\sum_{a}\left[(\alpha_a-\alpha_{n+1})X_a+\frac{{\beta_a}^2}{X_a}
\right]+\frac{{\beta_{n+1}}^2}{2(1-\sum_a X_a)}\ .
\end{eqnarray*}
Expanding in powers of the momenta, we see that
condition \rref{2} splits into
\begin{eqnarray}
\label{4}
&& d(L_{X_T}\theta_L-T dc_1)=0 \\
\label{5}
&& d(L_{X_V}\theta_L-V dc_1)=0\ .
\end{eqnarray}
\begin{rem}
\label{remrie}
As noticed in \cite{CraSarTho}, equation \rref{4} means that $L$ is a
symmetric conformal Killing tensor with respect to the usual Riemannian
metric of $S^n$, and implies that the torsion of $L$ vanishes. On the other
hand, equation \rref{5} can be written as
$d(L^*\, dV+c_1\, dV)=0$, which is a separability condition on the potential
$V$ appearing in the works of Benenti (see, e.g.,
\cite{Benjmp}). However, since our approach applies also to systems which
are defined on general symplectic manifolds (not necessarily cotangent
bundles), or, in other words, to Hamiltonians that are not quadratic in the
momenta, we will not use these results and we will solve directly equations
\rref{4} and \rref{5}. We also observe that a significant part of the
``Riemannian'' theory of separation of variables can be seen as a particular
case of the bi-Hamiltonian approach (see \cite{IMM,Gallipoli} and
\cite{Bl98}, where the Neumann system is also discussed).
\end{rem}

We start seeking a solution $L$ whose dependence on the coordinates $X_a$ is
affine: $L^a_b=\sum_{c} A_{bc}^aX_c +B_b^a$. Let us consider, for the sake of
simplicity, the case $n=2$. Condition \rref{4} gives
\[\begin{split}
&A_{12}^1=A_{22}^1=A_{11}^2=A_{21}^2=B_2^1=B_1^2=0\ ,\\
&A_{21}^1=A_{22}^2\ ,\quad A_{11}^1=A_{12}^2=A_{22}^2-B_1^1=B_2^2\ ,\end{split}
\]
so that we are left with the unknowns $A_{22}^2$, $B_1^1$, and $B_2^2$. Now,
condition \rref{5} is equivalent to
\[
A_{22}^2 (\alpha_2-\alpha_1)=(B_2^2-B_1^1)(\alpha_3-\alpha_2)\ ,
\]
which means that $A_{22}^2= c(\alpha_3-\alpha_2)$ and
$B_2^2-B_1^1=c(\alpha_2-\alpha_1)$ for some constant $c$. Thus
$B_2^2=c(\alpha_2+d)$ and
$B_1^1=c(\alpha_1+d)$, where $d$ is another constant, and the components of
$L$ are given by
\[
\left[\begin{array}{cc}
L^1_1 & L^1_2\\
L^2_1 & L^2_2\end{array}\right]=
c \left[\begin{array}{cc}
(\al_3-\al_1)X_1+\al_1+d & (\al_3-\al_2)X_1 \\
(\al_3-\al_1)X_2   & (\al_3-\al_2)X_2+\al_2+d \end{array}\right]\ .
\]
Since we are interested in the coordinates given by the eigenvalues of $L$,
we can set $c=1$ and $d=0$ without loss of generality.

Coming back to the general case, it is not difficult to check that the
1-form
\[
\theta_L=\sum_a \al_a Y_a\,dX_a+\left(\sum_b X_b Y_b\right)\sum_a
(\al_{n+1}-\al_a)\, dX_a\ ,
\]
corresponding to the $(1,1)$ tensor field given by
\begin{equation}\label{eq:gr1}
L^a_b=(\al_{n+1}-\al_b)X_a+\delta_b^a\al_a,
\end{equation}
satisfies conditions \rref{4} and \rref{5}.
Although these formulas define $L$ in coordinate patches,
it is not difficult to show that $L$ is globally defined
on the whole $S^n$.
Indeed, it is the restriction to $S^n$ of the tensor field $\hat L$ on
${\mathbb R}^{n+1}$ defined as
\[
{\hat L}\frac{\del}{\del x_i}=\al_i \frac{\del}{\del x_i}+
\frac{x_i}{r^4}\sum_{j,k} (\al_k-\al_j-\al_i){x_k}^2 x_j\,
\frac{\del}{\del x_j}
\]
where $r^2=\sum_i {x_i}^2$. In order to show that $\hat L$ restricts
to $S^n$ it is sufficient to check that ${\hat L}^\ast dr=0$, implying
that, at every point of $S^n$, the image of $\hat L$ is (contained in) the
tangent space to the sphere.

Summarizing, we have found a tensor field $L$ satisfying the separability
condition \rref{2}; thanks to the result of
\cite{CraSarTho} referred to in Remark \ref{remrie}, the torsion of $L$
vanishes. Thus the coordinates associated with $L$ are separated variables
for the NR system. Let us explicitly check that the eigenvalues of
$L$ coincide with the spheroconical coordinates.
To this end we find convenient to introduce
the following notations: let $\boldsymbol{\al}$ and
$\mathbf{X}$ denote the $n$--component vectors whose entries are,
respectively,
\[
\boldsymbol{\al}_b=\al_{n+1}-\al_b,\quad \mathbf{X}_b=X_b,
\qquad b=1,\ldots, n,
\]
and let $\mathbf{A}$ be the $n\times n$ diagonal matrix
of the parameters  $\al_a$. Then we can compactly write the matrix
form~\rref{eq:gr1} of the tensor field $L$ as
\begin{equation}
      \label{eq:gr2}
      L=\mathbf{A}+\mathbf{X}\otimes\boldsymbol{\al}.
\end{equation}
To compute the roots of $\text{det}(\la-L)$ we notice that the $L$ is a rank
$1$ perturbation of $\mathbf{A}$; hence we write
\begin{equation}
      \label{eq:gr3}
      \la-L=\big(\la-\mathbf{A}\big)
\cdot\left(\mathbf{1}-\mathbf{X}'(\la)\otimes\boldsymbol{\al}\right),
\end{equation}
where $\mathbf{X}'(\la)$ is the vector with entries
$\displaystyle{\frac{X_b}{\la-\al_b}} $.
Using the rank $1$ Aronszajn--Weinstein formula 
\[
\text{det}(\mathbf{1}+\mathbf{x}\otimes\mathbf{y})=1+\langle
\mathbf{y},\mathbf{x}\rangle,
\]
we arrive at
\begin{equation}
      \label{eq:gr4}
      \text{det}(\la-L)=\prod_{a=1}^n(\la-\al_a)\big(1-\sum_{b=1}^n\frac
{(\al_{n+1}-\al_b) X_b}{\la-\al_b}\big).
\end{equation}
Recalling the definitions $X_b={x_b}^2$, for $b=1,\ldots, n$, and the
constraint
$\sum_i {x_i}^2=1$, we can by means of elementary calculations conclude that
such an equation is equivalent to
\begin{equation}
      \label{eq:gr5}
      \text{det}(\la-L)=\prod_i (\la-\al_i)\sum_i \frac{x_i^2}{\la-\al_i},
\end{equation}
that is, the eigenvalues $\la_a$ of $L$ satisfy the equations
\begin{equation}
      \label{eq:gr6}
      \sum_i\frac{x_i^2}{\la-\al_i}=0 \>,\qquad \text { with }
      \sum_i x_i^2=1\ .
\end{equation}
These are the well-known defining relations for the spheroconical (or
elliptic-spherical) coordinates.

We close this section reporting, for the sake of completeness,
the well-known computation of the momenta $\mu_a$
conjugated to the $\la_a$. It is easily checked that the usual rule for
computing the residues gives
\begin{equation}
\label{6}
{x_i}^2=\frac{\prod_a (\al_i-\la_a)}{\prod_{j\ne i} (\al_i-\al_j)}\ .
\end{equation}
Then we have that
\[
{x_i}\,dx_i=-\frac12 \frac{\sum_a (\al_i-\la_a)\,d\la_a}
{\prod_{j\ne i} (\al_i-\al_j)}\ ,
\]
and, using again \rref{6}, that
\[
dx_i=\frac12 x_i\sum_a \frac{d\la_a}{\la_a-\al_i}\ .
\]
If $(x_1,\dots,x_{n+1},y_1,\dots,y_{n+1})\in {\mathbb R}^{2n+2}\supset
TS^n\simeq T^*S^n$, then the $\mu_a$ are given by
\[
\sum_a \mu_a\,d\la_a=\left(\sum_i y_i\,dx_i\right)_{|T^*S^n}=
\frac12 \sum_a
\left(\sum_i\frac{x_i y_i}{\la_a-\al_i}\right) d\la_a\ ,
\]
meaning that
\[
\mu_a=\frac12 \sum_i\frac{x_i y_i}{\la_a-\al_i}\ .
\]
Therefore, we can conclude that the separation variables $\la_a$ are the
solutions of
\[
\sum_i\frac{{x_i}^2}{\la-\al_i}=0\ ,
\]
while the conjugated momenta are given by $\mu_a=h(\la_a)$, with
\[
h(\la)=\frac12 \sum_i\frac{x_i y_i}{\la-\al_i}\ .
\]

\section{Integrals of motion and St\"ackel separability}\label{sec:3}

In the previous section we have found a tensor field $L$ on $S^n$ which
gives the separation coordinates of the NR system (i.e., the spheroconical
coordinates). Since $L$ satisfies the ``strong'' separability condition
\rref{2}, we know from Section 2 that:
\begin{enumerate}
\item There is an iterative method for constructing $n$ integrals of motion
in involution, $(H=H_1,H_2,\dots,H_n)$. (Of course, we have to take into
account that $T^*S^n$ is simply connected for $n\ge 2$.)
\item The NR system is St\"ackel-separable.
\end{enumerate}
The integrals of motion are given by
\begin{equation}
\label{7}
dH_{a+1}=N^* dH_a-c_a\,dH\ ,\qquad a=1,\dots,n-1\ ,
\end{equation}
where $\la^n-\sum_{a=1}^n c_a\la^{n-a}=\det (\la I-L)$. This defines  the
$H_a$ up to additive
constants. For example, in the case $n=2$ one finds
\[
\begin{aligned}
H_2 &=2(\al_1 {Y_2}^2 X_2+\al_2 {Y_1}^2 X_1)(X_1+X_2-1)-2\al_3
X_1 X_2(Y_2-Y_1)^2\\
&\ -\frac12 \left[\al_2(\al_1-\al_3) X_1+ \al_1(\al_2-\al_3) X_2
+\frac{\al_2{\be_1}^2}{X_1}+\frac{\al_1{\be_2}^2}{X_2}\right. \\
&\ \left. +(\al_3-\al_2){\be_1}^2 \frac{X_2}{X_1}
+(\al_3-\al_1){\be_2}^2 \frac{X_1}{X_2}
+{\be_3}^2\frac{\al_2 X_1+\al_1 X_2}{1-X_1-X_2}
\right]\ .
\end{aligned}
\]
Before showing that the $H_a$ coincide with the integrals of motion known in
the literature, let us consider the separability
{\em \`a la\/} St\"ackel of the NR system. It is guaranteed from the results
in Section 2 that
\begin{equation}
\label{7b}
\sum_{b=1}^n \la_a^{n-b} H_{b}=U_a(\la_a,\mu_a)\ ,\qquad a=1,\dots,n\ ,
\end{equation}
where the $U_a$ are polynomials.
\begin{rem}
It is easy to see that the polynomials $X_a=X_a(\la_1,\dots,\la_n)$ and
$Y_a=Y_a(\la_1,\dots,\la_n,\mu_1,\dots,\mu_n)$ are
invariant under the exchanges $(\la_b,\mu_b)\leftrightarrow (\la_c,\mu_c)$.
This entails that $U_b=U_c$.
\end{rem}

Next we want to compare the constants of motion defined by \rref{7}
with the spectral invariants of the Lax matrix
(see, e.g., \cite{Harnadlectures})
\[
N(\la)=\left[\begin{array}{cc}
-h(\la)+i k(\la)  &  e(\la)  \\
f(\la)      &   h(\la)+i k(\la)
\end{array}\right]\ ,
\]
where $h(\la)=\frac12\sum_i\frac{x_i y_i}{\la-\al_i}$ has already been
introduced, and
\[
k(\la)=\frac12 \sum_i\frac{\beta_i}{\la-\al_i}\ ,
e(\la)=-\frac12 \left(1+\sum_i\frac{{y_i}^2+\frac{{\beta_i}^2}
{{x_i}^2}}{\la-\al_i}\right)\ ,
f(\la)=\frac12 \sum_i\frac{{x_i}^2}{\la-\al_i}\ .
\]
The spectral invariants are the coefficients of the polynomial
\begin{equation}
\label{9}
P(\la)=a(\la)\det N(\la)=\frac14 \la^n +\sum_{a=1}^n P_a \la^{n-a}\ ,
\end{equation}
where $a(\la)=\prod_i (\la-\al_i)$ and the restriction to $T^*S^n$ has
been tacitly assumed.
In particular, $P_1=\frac12 H$, where $H$ is the NR Hamiltonian. Our
strategy to prove that $P_a=\frac12 H_a$ for all $a$ is to show that
\begin{equation}
\label{8}
N^*dP(\la_b)=\la_b\,dP(\la_b)\ ,\qquad b=1,\dots,n\ ,
\end{equation}
which implies that the $P_a$ satisfy
\[
N^* dP_a=dP_{a+1}+c_a\,dP_1\ ,\qquad a=1,\dots,n-1\ ,
\]
because the $\la_b$'s are the roots of \rref{2.7b}.
Since these relations coincide with the equations \rref{7} for the $H_a$,
and the starting points fulfill $P_1=\frac12 H_1$, we can conclude that
\begin{equation}
\label{10}
P_a=\frac12 H_a \qquad \mbox{for $a=1,\dots,n$.}
\end{equation}
To show that \rref{8} holds, we recall that $\mu_b=h(\la_b)$ and
$f(\la_b)=0$, so that \rref{9} entails
\begin{equation}
\label{9b}
P(\la_b)=a(\la_b)\left(-h(\la_b)^2-k(\la_b)^2-e(\la_b)f(\la_b)\right)
=-a(\la_b)\left({\mu_b}^2+k(\la_b)^2\right)\ .
\end{equation}
Then \rref{8} follows from the definition of DN coordinates.

Finally, from \rref{10} and \rref{9b} we obtain the separation relations for
the $H_a$,
\[
\sum_{a=1}^n H_a \la_b^{n-a}=-\frac12 \la_b^n -2a(\la_b)
\left({\mu_b}^2+k(\la_b)^2\right)\ ,
\]
i.e., the explicit form of the St\"ackel vector with components
$U_b$ appearing in \rref{7b}.

\vspace{0.5truecm}
\begin{flushleft}
\tiny
C. Bartocci and M. Pedroni:\\
Dipartimento di Matematica,
Universit\`a di Genova\\
Via Dodecaneso 35, I-16146 Genova, Italy\\
{\tt bartocci@dima.unige.it, pedroni@dima.unige.it}\\
\smallskip
G. Falqui:\\
SISSA, Via Beirut 2/4, I-34014 Trieste, Italy\\
{\tt falqui@sissa.it}
\end{flushleft}
\end{document}